\documentclass[prc,aps,amsfonts,amsmath,arydshln,superscriptaddress,floatfix,twocolumn]{revtex4}
\usepackage{graphicx,float} 
\usepackage{rotating}
\usepackage{braket}
\usepackage{blindtext}
\usepackage[colorlinks = true,
            linkcolor = blue,
            urlcolor  = blue,
            citecolor = blue,
            anchorcolor = blue]{hyperref}
\usepackage{ragged2e}

\newcommand{\di}{i}          

\newcommand{\tj}[6]{ \begin{pmatrix}
   #1 & #2 & #3 \\
   #4 & #5 & #6 
  \end{pmatrix}}

\begin{document}

\title{Cluster structures in $^{12}$C from global energy density functionals}

\author{P. Marevi\'c}
\affiliation{CEA,DAM,DIF, F-91297 Arpajon, France}
\affiliation{Institut de Physique Nucl\'eaire, Universit\'e Paris-Sud, IN2P3-CNRS, \\
Universit\'e Paris-Saclay, F-91406 Orsay Cedex, France}
\author{J.-P. Ebran}
\affiliation{CEA,DAM,DIF, F-91297 Arpajon, France}
\author{E. Khan}
\affiliation{Institut de Physique Nucl\'eaire, Universit\'e Paris-Sud, IN2P3-CNRS, \\
Universit\'e Paris-Saclay, F-91406 Orsay Cedex, France}
\author{T. Nik\v{s}i\'c}
\affiliation{Department of Physics, Faculty of Science, University of Zagreb, \\
Bijeni\v{c}ka c. 32, 10000 Zagreb, Croatia}
\author{D. Vretenar}
\affiliation{Department of Physics, Faculty of Science, University of Zagreb, \\
Bijeni\v{c}ka c. 32, 10000 Zagreb, Croatia}

\begin{abstract} 
 Spectroscopic properties of low-lying states and cluster structures in $^{12}$C are analyzed in a ``beyond mean-field framework" based on global energy density functionals (EDFs). To build symmetry-conserving collective states, axially-symmetric and reflection-asymmetric solutions of the relativistic Hartree-Bogoliubov equations are first projected onto good values of angular momentum, particle number, and parity. Configuration mixing is implemented using the generator coordinate method formalism. It is shown that such a global microscopic approach, based on a relativistic EDF, can account for the main spectroscopic features of $^{12}$C, including the ground-state and linear-chain bands as well as, to a certain approximation, the excitation energy of the Hoyle state. The calculated form factors reproduce reasonably well the available experimental values, and display an accuracy comparable to that of dedicated microscopic cluster models. 
\end{abstract}

\date{\today}

\maketitle

\section{Introduction}

The formation of clusters, as transitional states between the quantum liquid and solid phases of finite systems, is a common feature in light atomic nuclei \cite{beck10,beck12,beck14,horiuchi12,freer17}. Particularly favorable conditions for the appearance of cluster structures are found in light self-conjugate nuclei, where various exotic configurations have been predicted and some observed. Probably one of the most prominent examples is the self-conjugate nucleus $^{12}$C, in which axial oblate, triangular, linear $3\alpha$, and bent arm configurations are all predicted to coexist at low excitation energies \cite{tohsaki01,kanada07,chernykh07,neff04,funaki15,fukuoka13,ren18,tohsaki17}. New reorientation-effect measurements have very recently confirmed the pronounced oblate deformation in the ground-band $2_1^+$ state of $^{12}$C \cite{kumar18}. First experimental evidence of triangular $\mathcal{D}_{3h}$ symmetry in $^{12}$C was reported by Mar\'in-Lamb\'arri \emph{et al.}, with $K=0$ and $K=3$ bands reportedly merging to form a single rotational band built on triaxial ground state \cite{lambarri14}. Exceptional results on the structure of the well known $0_2^+$ (Hoyle) state have been simultaneously reported by two independent groups \cite{dellaquila17,smith17}, indicating that this state predominantly decays by intermediate emission of an $\alpha$-particle. In parallel with experimental advances, a number of theoretical methods have been employed to study the variety of shapes in $^{12}$C, including the antisymmetrized molecular dynamics (AMD) model \cite{kanada07}, the fermionic molecular dynamics (FMD) model \cite{neff04,chernykh07}, the Tohsaki-Horiuchi-Schuck-R\"{o}pke (THSR) wave function model \cite{tohsaki01,funaki15}, configuration-mixing Skyrme calculations \cite{fukuoka13}, and the cranking relativistic mean-field theory \cite{ren18}. Both the AMD and FMD calculations predict the Hoyle state to be a weakly-interacting assembly of $^{8}$Be$+\alpha$ configurations \cite{kanada07,chernykh07}, while the THSR model describes the Hoyle state in terms of a condensate of $\alpha$-particles \cite{tohsaki01}. In addition, $3\alpha$ linear chain structures were predicted to occur in higher $0^+$ states, even though the stability of these configurations against bending remains an open question \cite{kanada07,chernykh07,ren18}.

The framework of relativistic energy density functionals (EDFs) \cite{VALR.05,niksic11} has been extensively used in studies of cluster structures on both the mean-field level \cite{arumugam05,maruhn06,ebran12,ebran14a,ebran14b,ebran18} and, to a lesser extent, using ``beyond mean-field'' \cite{zhou16,marevic18} methods. Particularly interesting results on the origins of nuclear clustering have been reported in Ref. \cite{ebran12}, where the appearance of pronounced cluster structures has been linked to the  underlying single-nucleon potential. However, in order to carry out a quantitative analysis that can directly be compared to experiment, it is necessary to extend  the simple mean-field picture by taking into account collective correlations related to restoration of broken symmetries and configuration mixing. Symmetry-conserving EDF-based methods have been used to model a variety of structure phenomena over the entire nuclide chart \cite{bender03,niksic11,egido16}. One of the major advantages of using such an approach in studies of clusters is that it is not necessary to \emph{a priori} assume the existence of localized structures in the model space. In fact, the EDF framework includes both the quantum-liquid and cluster aspects of nuclear systems, and clusterization may eventually occur as a result of the self-consistent approach on mean-field level and/or the subsequent restoration of symmetries and configuration mixing. In addition, and this is important, the parameters of global functionals are typically adjusted to data on medium-heavy and heavy nuclei and the effective interaction itself does not bear any information specific to light systems, or to cluster states that one aims to describe. 

In this work we employ the framework of symmetry-conserving relativistic EDF to study cluster structures in positive-parity states of $^{12}$C isotope. Axially-symmetric and reflection-asymmetric reference states are generated as solutions to the relativistic Hartree-Bogoliubov (RHB) equations. These configurations are then projected onto good values of angular momenta, particle number, and parity, before being mixed using the generator coordinate method (GCM). This paper is organised as follows. A brief outline of the theoretical method is presented in Section \ref{sec:theory}. In Section \ref{sec:results} we demonstrate how this global model can account for the main spectroscopic features of $^{12}$C, including the description of both the ground-state and linear-chain bands, as well as the excitation energy of the Hoyle state. Furthermore, the theoretical  elastic and inelastic form factors are shown to reproduce experimental values, and exhibit an accuracy that can compete with dedicated microscopic cluster models. Finally, Section \ref{sec:conclusion} briefly summarizes the main results of this work.

\section{Theoretical Framework}
\label{sec:theory}

The EDF-based symmetry conserving and configuration mixing approach provides a global method that can be applied to studies of structure phenomena over the entire nuclide chart \cite{bender03,niksic11,egido16}. In practical implementations this framework essentially presents a two-step process. In the first step (single-reference EDF) a number of symmetries of the nuclear Hamiltonian are broken at the self-consistent mean-field level. This approach provides an approximate description of nuclear ground states in terms of symmetry-breaking many-body wave functions. Bulk properties of atomic nuclei (such as binding energies, charge radii, etc.) can be analyzed at this level. In the second step (multi-reference EDF), previously broken symmetries of many-body states are recovered and the resulting configurations are further mixed to construct collective states with good quantum numbers. Going beyond the simple mean-field picture, one is able to take into account additional collective correlations and thereby describe a nucleus in the laboratory frame, including spectroscopic properties. The GCM ansatz for the symmetry-conserving collective state $\ket{\Psi^{J;NZ;\pi}_{\alpha}}$ reads \cite{RS.80}:

\begin{equation}
\ket{\Psi^{J;NZ;\pi}_{\alpha}} = \sum_{q_i} \sum_K f^{J;NZ;\pi}_{\alpha}(q_i) \hat{P}^{J}_{MK} \hat{P}^N \hat{P}^Z \hat{P}^{\pi} \ket{\Phi(q_i)},
\label{eq:gcmstate}
\end{equation}

\noindent with the first summation running over a discretized set of quadrupole and octupole deformations $\{ q_i \} \equiv \{ {\beta_2}_i, {\beta_3}_i \}$. Here we define the dimensionless deformation parameters $\beta_{\lambda} = 4 \pi q_{\lambda 0}/3 A R^2$, where $q_{\lambda 0}$ and $R=r_0 A^{1/3}$ denote the mass multipole moment and  nuclear radius, respectively. Furthermore, $\hat{P}^{J}_{MK}$ is the angular momentum projection operator:

\begin{equation}
\hat{P}^{J}_{MK} = \frac{2J+1}{8\pi^2} \int \,d\Omega D^{J*}_{MK} (\Omega) \hat{R}(\Omega),
\label{eq:projector_angular}
\end{equation}

\noindent where the integral is over the three Euler angles $\Omega = (\alpha, \beta, \gamma)$, $D^{J}_{MK}$ is Wigner $D$ matrix \cite{varshalovich88}, and $\hat{R}=e^{-i\alpha\hat{J}_z}e^{-i\beta\hat{J}_y}e^{-i\gamma\hat{J}_z}$ is the rotation operator. The projection operators onto neutron $\hat{P}^N$, and proton number $\hat{P}^Z$ read:

\begin{equation}
\hat{P}^{N_{\tau}} = \frac{1}{2\pi} \int_0^{2 \pi} \,d\varphi  e^{i (\hat{N}_{\tau} - N_{\tau}) \varphi}, \quad \tau = N, Z.
\end{equation}

\noindent $\hat{N}_{\tau}$ corresponds to the particle number operator and $N_{\tau}$ is the desired number of nucleons in each isospin channel $\tau$. We note that because of symmetry the integration interval can be reduced to $[0, \pi]$ for even-even nuclei. In addition, projection onto good parity is performed by choosing a basis that is closed under parity transformation, that is, by ensuring that for each $(\beta_2, \beta_3)$ state the basis always includes the corresponding $(\beta_2, -\beta_3)$ state. The parity quantum number is then restored by the subsequent configuration mixing.

The RHB states $\ket{\Phi(q_i)}$ are obtained from deformation-constrained self-consistent mean-field calculations using the relativistic point-coupling functional DD-PC1 \cite{niksic08} in the particle-hole channel, and a separable pairing force in the particle-particle channel \cite{duguet04,tian09}. The RHB basis of the present calculation  comprises a wide range of both quadrupole and octupole deformations: $\beta_2 \in [-1.2, 3.6]$ and $\beta_3 \in [-3.5, 3.5]$, with mesh sizes $\Delta \beta_2 = 0.4$, $\Delta \beta_2 = 0.6$, and $\Delta \beta_3 = 0.7$ in the oblate, prolate, and octupole directions, respectively. To ensure a proper convergence, the RHB states are expanded in a basis of the axially-symmetric harmonic oscillator \cite{gambhir90} with $N_{\text{sh}} = 10$ ($N_{\text{sh}} = 11$) oscillator shells for the large (small) component of the Dirac single-nucleon spinor. Furthermore, the axial, time-reversal, and simplex symmetry of RHB states are imposed. In particular, axial symmetry reduces the computational task considerably, as integrals over the Euler angles $\alpha$ and $\gamma$ in Eq. (\ref{eq:projector_angular}) can be carried out analytically. Projection integrals over the gauge angle $\varphi$ are performed using the standard Fomenko expansion \cite{fomenko70}. The corresponding number of integration points in the Euler angle $\beta$ and gauge angle $\varphi$ are $N_{\beta} = 27$ and $N_{\varphi} = 9$, respectively. The weight functions $f^{J;NZ;\pi}_{\alpha}(q_i)$ of Eq. (\ref{eq:gcmstate}) are determined by solving the Hill-Wheeler-Griffin (HWG) equation \cite{hwg}:

\begin{equation}
\begin{aligned}
& \sum_{q_j} \mathcal{H}^{J;NZ;\pi}(q_i, q_j) f_{\alpha}^{J;NZ;\pi}(q_j) = \\ & E_{\alpha}^{J;NZ;\pi} \sum_{q_j} \mathcal{N}^{J;NZ;\pi}(q_i, q_j)   f_{\alpha}^{J;NZ;\pi}(q_j) .
\label{eq:hwg}
\end{aligned}
\end{equation}

\noindent The norm overlap kernel $\mathcal{N}^{J;NZ;\pi}(q_i, q_j)$ and the Hamiltonian kernel $\mathcal{H}^{J;NZ;\pi}(q_i, q_j)$ are given by the generic expression:

\begin{widetext}
\[
\mathcal{O}^{J;NZ;\pi}(q_i, q_j) = \frac{2J+1}{2}\delta_{M0}\delta_{K0} \int_0^{\pi} \,d\beta \sin\beta d^{J*}_{00}(\beta) \frac{1}{N_{\varphi}^2} \sum_{l_N, l_Z=1}^{N_{\varphi}} e^{-\di \varphi_{l_N} N_0} e^{-\di \varphi_{l_Z} Z_0} \braket{\Phi(q_i)|\hat{O} e^{-\di \beta \hat{J}_x} e^{\di \varphi_{l_N} \hat{N}} e^{\di \varphi_{l_Z} \hat{Z}} \hat{P}^{\pi} |\Phi(q_j)},
\]
\end{widetext}

\noindent with $\hat{O} = \hat{1}, \hat{H}$ for the norm overlap kernel and the Hamiltonian kernel, respectively. We note that here the Hamiltonian kernel is calculated using the mixed density prescription \cite{robledo10}. The numerical solution of the HWG equation for a given angular momentum and parity yields the lowest collective state $(\alpha = 1)$, as well as excited states $(\alpha = 2, 3, ...)$. Additionally, one can use the weights $f^{J;NZ;\pi}_{\alpha}(q_i)$ to define another set of functions:

\begin{equation}
g^{J;NZ;\pi}_{\alpha}(q_i) = \sum_{q_j} (\mathcal{N}^{J;NZ;\pi})^{1/2}(q_i, q_j) f_{\alpha}^{J;NZ;\pi} (q_j). 
\end{equation}

\noindent Since $g^{J;NZ;\pi}_{\alpha}(q_i)$ are orthonormal, they are interpreted as collective wave functions of the variables $q_i$. Furthermore, even though they are not observables themselves, the collective wave functions explicitly manifest shape fluctuations in both the quadrupole and octupole directions and can be used to calculate various observables, such as spectroscopic quadrupole moments and electromagnetic transition rates \cite{rodriguez02}. In particular, the electric spectroscopic quadrupole moment of the collective state $\ket{\Psi^{J;NZ;\pi}_{\alpha}}$ can be computed from the expression

\begin{align}
\begin{split}
Q_2^{\text{spec}}&(J^{\pi}_{\alpha}) = e \sqrt{\frac{16 \pi}{5}} \tj{J}{2}{J}{J}{0}{-J} \\ & \times \sum_{q_i q_f}  f_{\alpha}^{J;NZ;\pi *}(q_f) \braket{J q_f || \hat{Q}_{2} || J q_i} f_{\alpha}^{J;NZ;\pi}(q_i). 
\end{split}
\end{align}

\noindent The reduced electric multipole transition probability for
a transition between the initial $\ket{\Psi^{J_i;NZ;\pi_i}_{\alpha_i}}$ and final state $\ket{\Psi^{J_f;NZ;\pi_f}_{\alpha_f}}$ reads:

\begin{align}
\begin{split}
&B(E\lambda; J^{\pi_i}_{\alpha_i} \rightarrow J^{\pi_f}_{\alpha_f})  = \frac{e^2}{2 J_i + 1} \\ & \times {\Big | } \sum_{q_i q_f} f_{\alpha_f}^{J_f;NZ;\pi_f *}(q_f) \braket{J_f q_f || \hat{Q}_{\lambda} || J_i q_i} f_{\alpha_i}^{J_i;NZ;\pi_i}(q_i) {\Big |}^2.
\end{split}
\end{align}

\noindent Furthermore, utilizing techniques recently developed for Skyrme-based EDF calculations \cite{yao15}, collective wave functions can also be used to compute elastic and inelastic form factors for electron scattering. In the plane-wave Born approximation the longitudinal Coulomb form factor $F_L(q)$ for a transition from the initial state $\ket{\Psi^{J_i;NZ;\pi_i}_{\alpha_i}}$ to the final state $\ket{\Psi^{J_f;NZ;\pi_f}_{\alpha_f}}$ reads \cite{yao15}:

\begin{equation}
F_L(q) = \frac{\sqrt{4 \pi}}{Z} \int_0^{\infty} \,dr r^2 \rho^{J_f \alpha_f}_{J_i \alpha_i, L}(r) j_L(qr),
\label{eq:formfactor}
\end{equation}

\noindent where $q$ denotes the momentum transfer for angular momentum $L$, $j_L(qr)$ is the spherical Bessel function of the first kind, and $\rho^{J_f \alpha_f}_{J_i \alpha_i, L}(r)$ are reduced transition densities of protons. The latter can be computed from:

\begin{equation}
\begin{aligned}
\rho^{J_f \alpha_f}_{J_i \alpha_i, L} (r) &= (-1)^{J_f - J_i} \frac{2J_f+1}{2J_i+1} \sum_K \braket{J_f 0 L K|J_i K} \\ & \times \int \,d \mathbf{\hat{r}} \rho^{J_f J_i K 0}_{\alpha_f \alpha_i} (\mathbf{r}) Y_{LK}^{*}(\hat{\mathbf{r}}),
\end{aligned}
\label{eq:redtrans}
\end{equation}

\noindent where $\rho^{J_f J_i K 0}_{\alpha_f \alpha_i} (\mathbf{r})$ stands for the pseudo-GCM density as defined in Ref. \cite{yao15}. The pseudo-GCM density does not represent an observable, rather it encapsulates all the information related to the solution of HWG equation. More details on calculations of pseudo-GCM densities can be found in Ref. \cite{yao15}. Here we note that, to account for the spurious center-of-mass motion, a simple correction is introduced by folding the calculated form factors of Eq. (\ref{eq:formfactor}) with $G_{\text{cm}}(q) = \text{exp}[q^2 b^2/(4A)]$, where $b = \sqrt{\hbar/m\omega}$ denotes the oscillator length \cite{brown83}. It should be emphasized  that, since all quantities are calculated in the full configuration space, there is no need for effective charges and thus $e$ denotes the bare proton charge.

\section{Cluster structures in $^{12}$C}
\label{sec:results}

\subsection{Deformation Energy Maps}

\begin{figure*}[!t]
\includegraphics[width=0.92\textwidth]{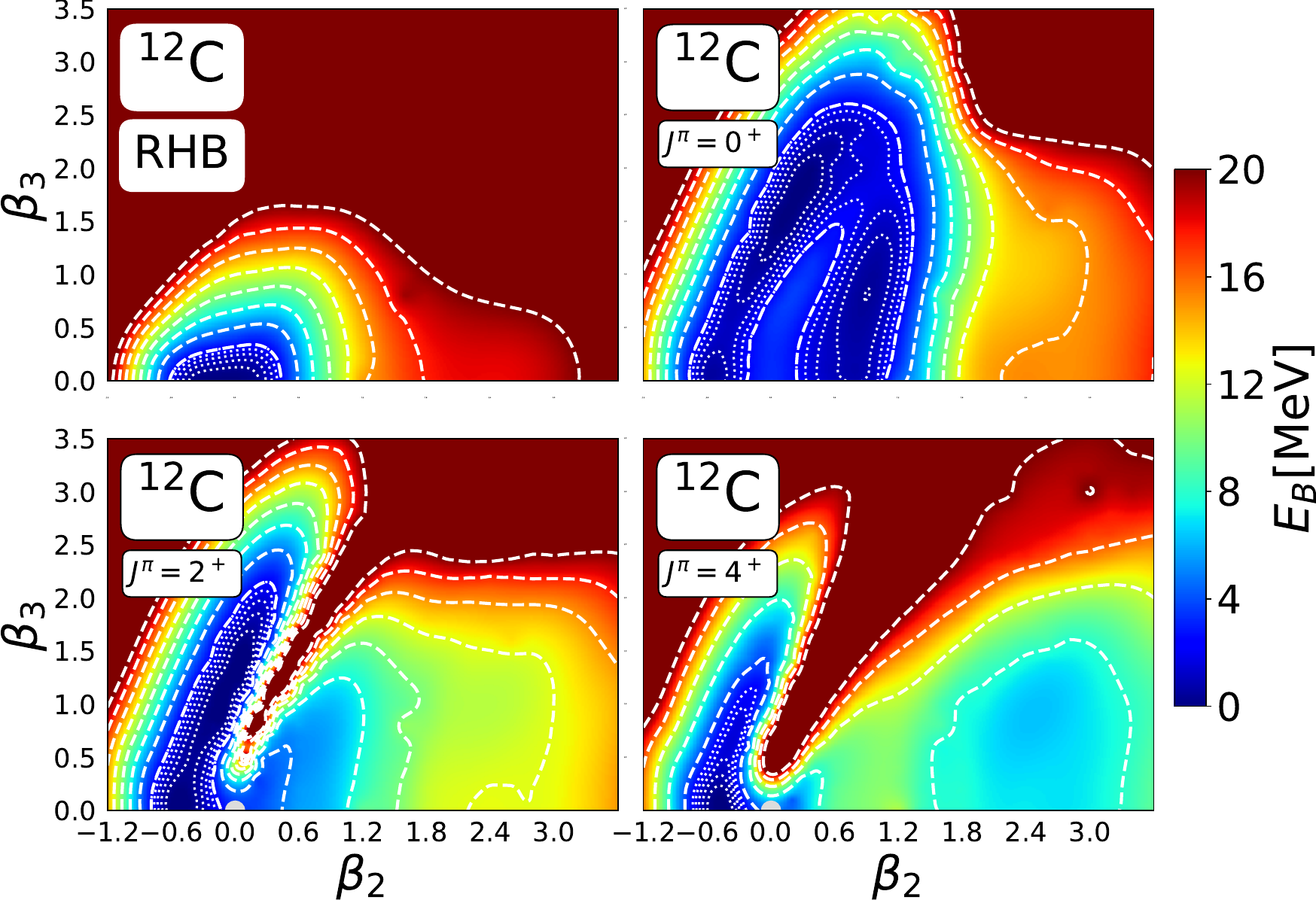}
\caption{Deformation energy surfaces of $^{12}$C in the $\beta_2$ - $\beta_3$ plane. In addition to the self-consistent mean-field RHB binding energies (upper left panel), the angular momentum-, particle number-, and parity-projected energy surfaces are shown for spin-parity values $J^{\pi} = 0^+ , 2^+, 4 ^+$. For each surface, energies are normalized with respect to the corresponding absolute minimum. Contour lines are separated by $2.5$ MeV (dashed lines) and $0.5$ MeV (dotted lines).}
\label{fig:PROJ}
\end{figure*}

Our analysis of cluster structures in $^{12}$C starts with a microscopic self-consistent mean-field RHB calculation. The upper left panel of Figure \ref{fig:PROJ} displays 
the deformation energy surface in the $\beta_2$ - $\beta_3$ plane. Although the absolute minimum of the RHB energy surface is found for the spherical $(\beta_2 = 0, \beta_3 = 0)$ configuration, we note that the surface is rather soft for the following range of deformations: $\beta_2 \approx [-0.6, 0.6]$ and $|\beta_3| \approx [0.0, 0.5]$. To illustrate the effect of symmetry restoration on the topology of the RHB energy, the other panels in Fig. \ref{fig:PROJ} show the corresponding angular momentum-, particle number-, and parity- projected energy surfaces for spin-parity values $J^{\pi} = 0^+, 2^+ , 4^+$. Already the $J^{\pi} = 0^+$ energy surface exhibits the dramatic impact of symmetry restoration. In particular, the $J^{\pi} = 0^+$ surface appears significantly softer for a wide range of deformations in comparison to the RHB mean-field energy surface. For higher values of angular momentum the minimum on the oblate side becomes much more pronounced, while intermediate prolate deformations become increasingly less  favored. 

\begin{figure}[!t]
\includegraphics[width=0.49\textwidth, trim={2cm 0 5cm 0},clip]{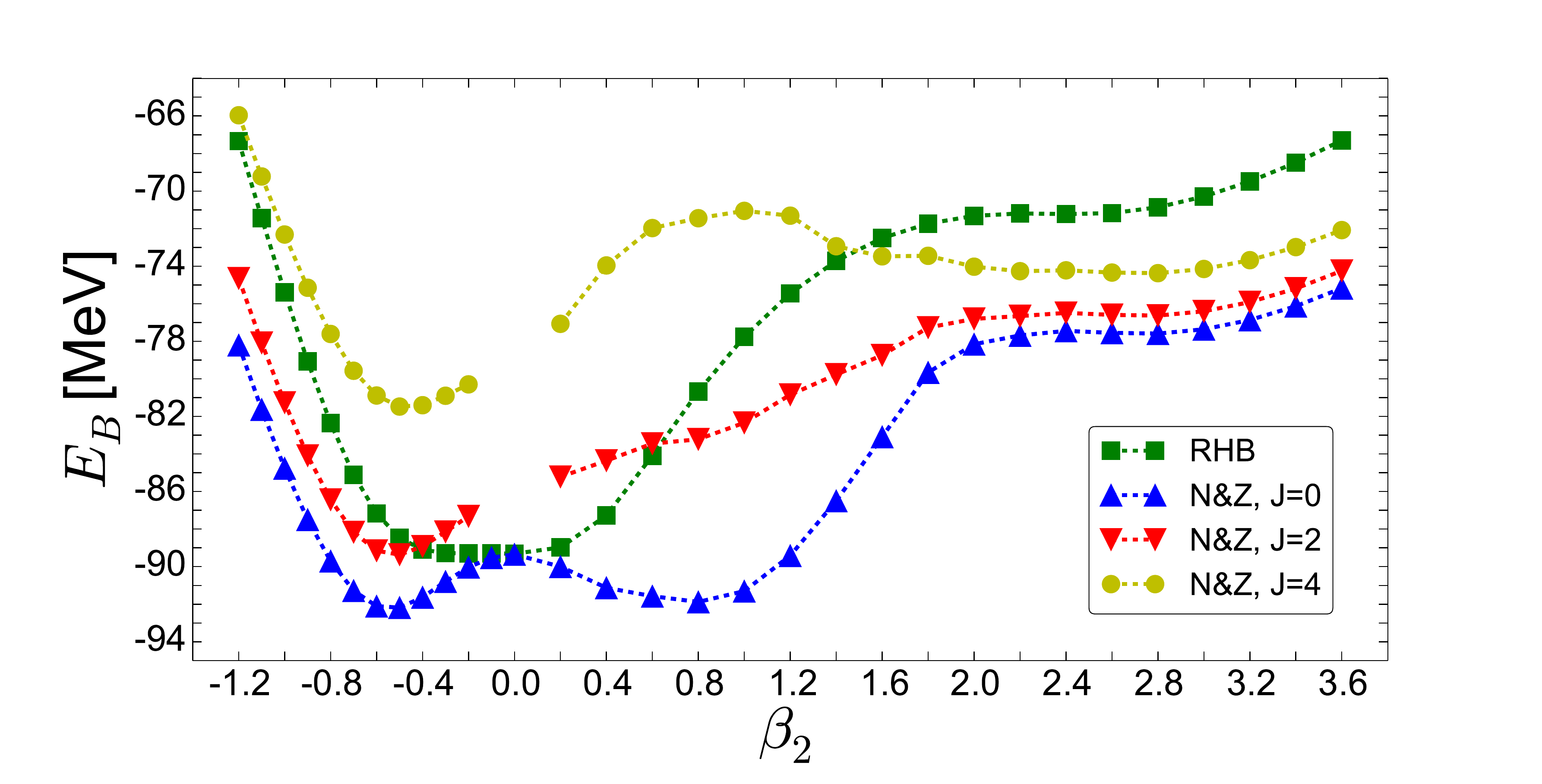}
\caption{Energy curves of $^{12}$C as functions of the axial quadrupole deformation $\beta_2$ for parity-conserving ($\beta_3 = 0$) configurations. In addition to the self-consistent mean-field RHB binding energies (squares), we display the angular momentum- and particle number-projected curves for spin-parity values $J^{\pi} = 0^+ , 2^+ , 4^+$.}
\label{fig:PROJ1D}
\end{figure}

\indent This trend is even more apparent in Figure \ref{fig:PROJ1D}, where we make a cut along the parity-conserving ($\beta_3 = 0$) line of Fig. \ref{fig:PROJ} and show the energy curves as functions of the axial quadrupole deformation $\beta_2$. As noted before, the RHB energy curve is rather flat around the spherical minimum. The only indication of possible cluster formation is found at very large prolate deformations, where a shoulder in the binding energy curve occurs. In contrast, the energy curve that corresponds to the symmetry-restored $J^{\pi} = 0^+$ states exhibits two nearly-degenerate minima: the lower one at $\beta_2 \approx -0.5$, and the prolate minimum at $\beta_2 \approx 0.8$. We note that the binding energy of the symmetry-restored oblate minimum is rather close to the experimental value: $E_B = -92.16$ MeV. This is consistent with the fact that the collective $0_1^+$ state is expected to correspond to the band-head of an oblate-deformed rotational band. Furthermore, the shoulder at large quadrupole deformations is preserved for the $J^{\pi} = 0^+$ projected energy curve and it is additionally lowered by about $5$ MeV. The $J^{\pi} = 2^+$ projected curve preserves both the oblate minimum and the shoulder at large quadrupole deformations, while the former $J^{\pi} = 0^+$ local minimum at intermediate quadrupole deformation transforms into a shoulder. Finally, a barrier occurs at intermediate prolate deformations for the $J^{\pi} = 4^+$ state.

\indent The analysis of Figs.~\ref{fig:PROJ} and \ref{fig:PROJ1D} illustrates how the symmetry-restored potential energy maps already encapsulate the variety of shapes of $^{12}$C. Nevertheless, it is only by performing configuration mixing, that is, including collective correlations related to both quadrupole and octupole shape fluctuations, that one obtains a quantitative description of $^{12}$C spectroscopic properties.

\subsection{Spectroscopy of Collective States}

\begin{figure}[!t]
\includegraphics[scale=0.35]{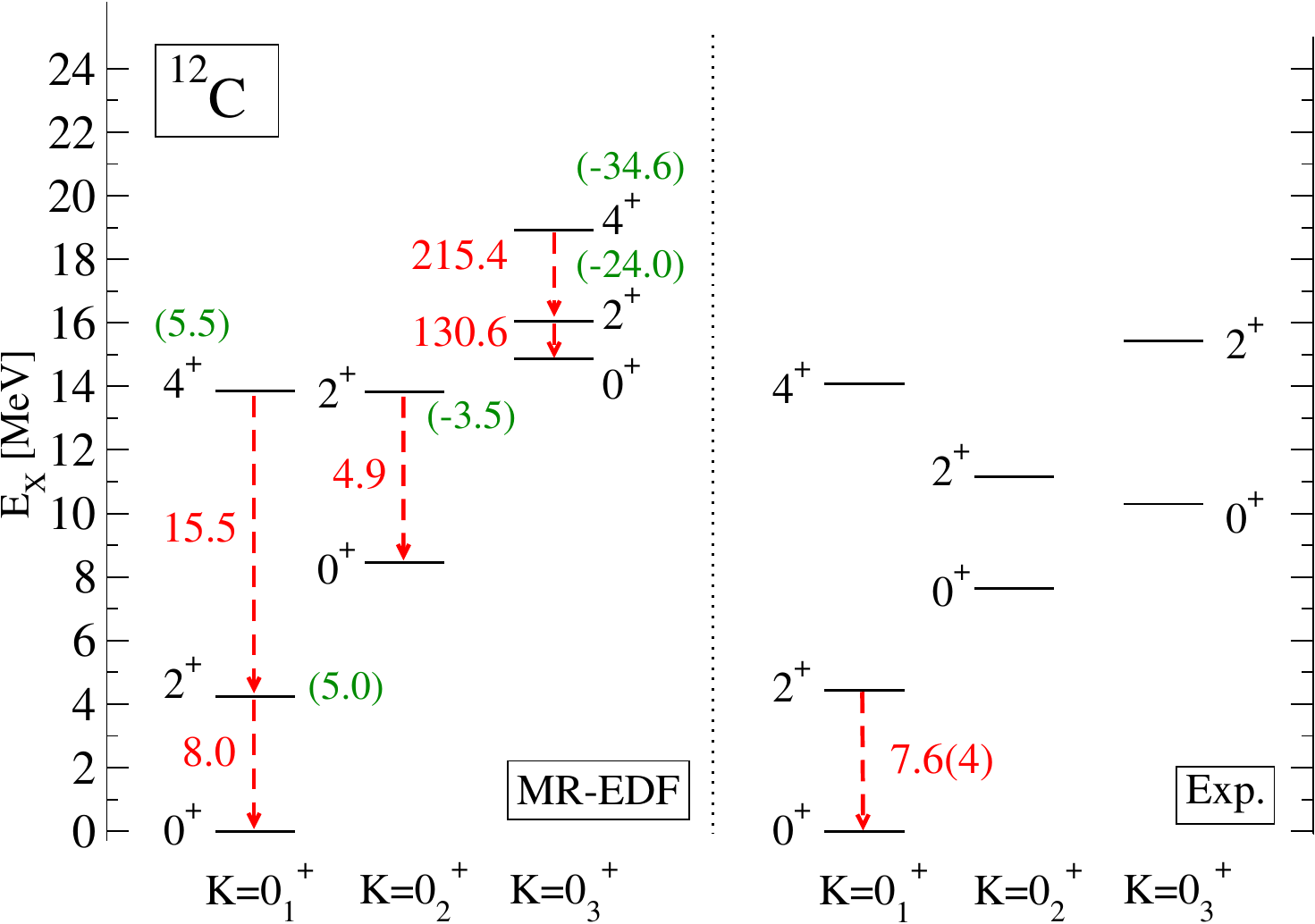}
\caption{The calculated low-energy positive-parity excitation spectrum of $^{12}$C compared to the available data \cite{nndc}. Intraband $B(E2)$ transition strengths (red color, in $e^2$fm$^4$) and spectroscopic quadrupole moments (green color, in $e$~fm$^2$) are also shown. See text for more details.}
\label{fig:spectrum}
\end{figure}

In the next step, $72$ symmetry-restored configurations were mixed for each angular momentum using the GCM as described in the previous section. RHB configurations with binding energies much higher than energy of the equilibrium configuration ($30$ MeV and higher) have been excluded from the basis. Of course, we have verified that this choice of the basis does not affect the calculated low-energy spectroscopic properties. Figure \ref{fig:spectrum} displays the calculated low-lying positive-parity spectrum in comparison to the available data. The intraband $B(E2)$ transition strengths and spectroscopic quadrupole moments of collective states are also shown. The theoretical excitation energies of the $2_1^+$ ($4.3$ MeV) and $4_1^+$ ($13.9$ MeV) states are only slightly lower than the corresponding experimental values, $4.44$ MeV and $14.08$ MeV, respectively \cite{kelley17}. In addition, their ratio $E(4_1^+)/E(2_1^+) = 3.23$ is in very good agreement with the experimental value of $3.17$, pointing to the rotational character of the ground-state band. Furthermore, the calculated $E2$ transition strength from the $2_1^+$ state to the ground state, $B(E2; 2_1^+ \rightarrow 0_1^+) = 8.0$ $e^2$fm$^4$, reproduces the experimental value. We also note that the theoretical spectroscopic quadrupole moment of the $2_1^+$ state, $Q_{\text{spec}} (2_1^+) = 5.0$ $e$fm$^2$, is predicted within the error bar of the very recent experiment: $Q^{\text{exp}}_{\text{spec}} (2_1^+) = (7.1 \pm 2.5)$ $e$fm$^2$ \cite{kumar18}. It is therefore interesting to point out that, in this specific instance, our global EDF-based approach provides a level of agreement with data that is comparable to state-of-the-art \textit{ab initio} models \cite{kumar18}. Finally, the calculated $B(E2; 4_1^+ \rightarrow 2_1^+) = 15.5$ $e^2$fm$^4$ agrees with the AMD prediction $B(E2; 4_1^+ \rightarrow 2_1^+) = 16$ $e^2$fm$^4$ \cite{kanada07}. This transition has yet to be measured, and its strength would definitely confirm the rotational character of the ground-state band.

\indent The $4^+$ state calculated at $\approx 19$ MeV is characterized by a pronounced prolate deformation, a feature shared by the $0_3^+$ and $2_3^+$ collective states. In addition, it exhibits a very strong $E2$ transition strength to the $2_3^+$ state. Therefore, it is assigned to the $K^{\pi} = 0_3^+$ band and denoted as $4_3^+$ further on. Both the very large $E2$ intraband transitions and the value of the ratio $(E(4_3^+)-E(0_3^+))/(E(2_3^+)-E(0_3^+))=3.45$ characterize the rotational nature of this band. In particular, the calculated value of $B(E2; 2_3^+ \rightarrow 0_3^+) = 130.6$ $e^2$fm$^4$ is somewhat larger than the one reported in Ref. \cite{kanada07}, but still of the same order of magnitude. Additionally, the huge spectroscopic quadrupole moments of $4_3^+$ and $2_3^+$ states support the interpretation of this band as a $3\alpha$ linear chain.

\indent In the present study the $0_2^+$ (Hoyle) state is calculated at an excitation energy that is only about $800$ keV above the experimental value. However, the $E2$ transition strength from the corresponding $2_2^+$ state is about an order of magnitude smaller than the one obtained in the AMD \cite{kanada07} and THSR \cite{funaki15} calculations. Even though there are currently no available data, it is likely that our calculation actually underestimates the true value for this transition strength. The reason is at least two-fold. Firstly, the AMD and THSR models consistently predict triaxial configurations as the dominant contribution to the Hoyle state intrinsic density. These configurations are not included in the model space of the present study, and an extended analysis that allows for additional breaking of axial symmetry is necessary for a quantitative comparison. In addition, the asymptotic behavior of three weakly-interacting $\alpha$ particles is notoriously complicated to describe using the harmonic oscillator basis \cite{navratil07,neff12}. Therefore, it remains an open question whether self-consistent models based on harmonic oscillator bases, even including triaxial shapes, will be able to capture all the details of the Hoyle state density profile. Nevertheless, we note that the calculated transition strength from the $2_1^+$ state to the Hoyle state, $B(E2; 2_1^+ \rightarrow 0_2^+) = 1.7$ $e^2$fm$^4$ compares favorably to the experimental value $2.6 \pm 0.4$ $e^2$fm$^4$ and to the AMD prediction ($5.1$ $e^2$fm$^4$)  \cite{kanada07}. In addition, even the transition strength from the $2_2^+$ state to the ground state, $B(E2; 2_2^+ \rightarrow 0_1^+) = 1.3$ $e^2$fm$^4$, is in qualitative agreement with THSR prediction ($2.0-2.5$ $e^2$fm$^4$) \cite{funaki15}, when compared to the experimental value ($0.73 \pm 0.13$ $e^2$fm$^4$).

\indent For completeness, we note that the measured low-lying spectrum of $^{12}$C also includes a number of negative-parity states. Most notably, the $3_1^-$ state at $9.6$ MeV is considered a candidate for the $K^{\pi} = 3^-$ band head \cite{freer17}, a suggestion that is supported by, for example, AMD \cite{kanada07} and $3\alpha$GCM calculations \cite{uegaki77}. On the other hand, recent measurements suggest possible merging of this band with the $K^{\pi}=0^+$ band by forming a single rotational band built on triaxial ground state \cite{lambarri14}. Furthermore, in the same work, the $1_1^-$ state at $10.84$ MeV was interpreted as a band head of the vibrational bending mode whose lowest-lying rotational excitations consist of nearly degenerate parity doublets of $2^{\pm}$ and $3^{\pm}$ states. However, since only the corresponding $2^-$ excitation has been observed so far, this state could also represent a $K^{\pi}=1^-$ band head \cite{kanada07,uegaki77}. In any case, the imposed symmetries of the present study restrict our access to $K^{\pi} = 0^{\pm}$ bands only and we do not account for any of these two states. The lowest $1^-$ and $3^-$ states in our calculation are found at $13.2$ and $19.6$ MeV, respectively, and a rather large quadrupole transition strength between them, $B(E2; 3_1^- \rightarrow 1_1^-) = 3.6$ $e^2$fm$^4$, suggests they are members of the same band.

\subsection{Collective States in the Intrinsic Frame}

\begin{figure*}
\includegraphics[scale=0.32]{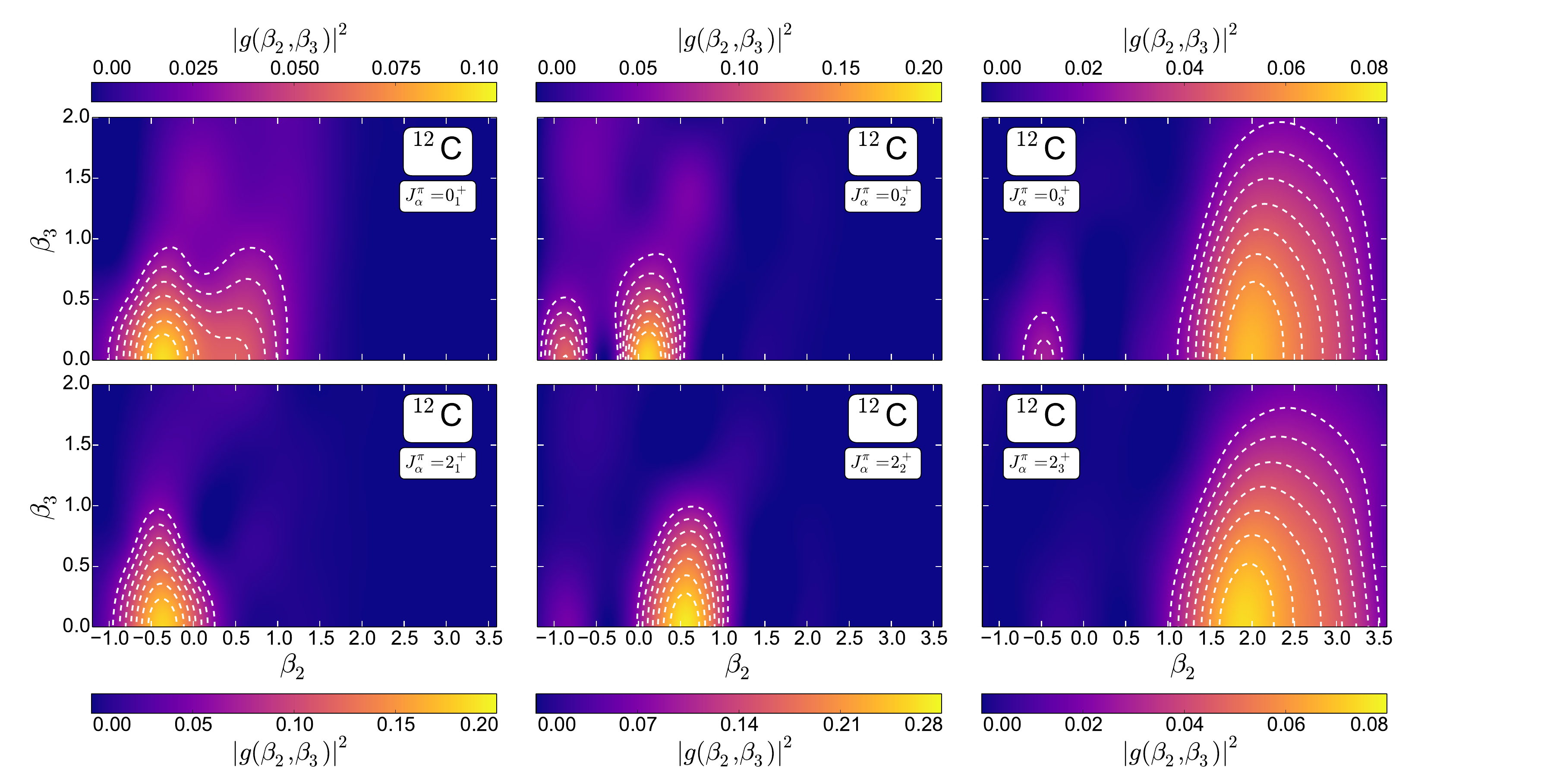}
\caption{Amplitudes of collective wave functions squared $|g(\beta_2, \beta_3)|^2$ for the low-energy levels of $^{12}$C. Dashed contours in the $\beta_2 - \beta_3$ plane successively denote a $10\%$ decrease starting from the largest value of the amplitude.}
\label{fig:waves}
\end{figure*}

\begin{figure*}[!t]
\includegraphics[scale=0.38]{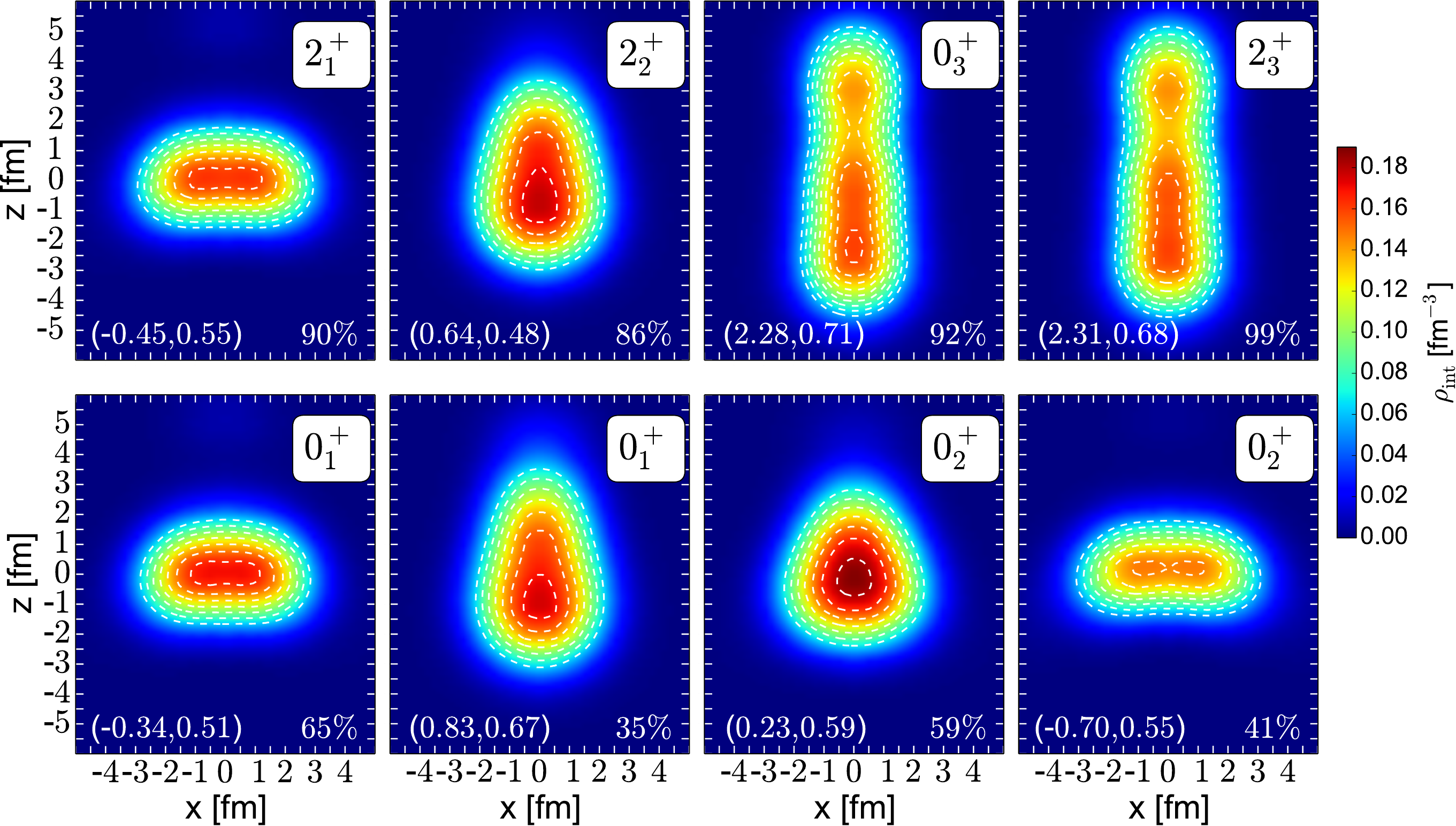}
\caption{Characteristic intrinsic nucleon densities of the first three $0^+$ and $2^+$ collective states in $^{12}$C. The corresponding average deformation parameters $(\beta_2, \beta_3)$, as well as the respective contributions of prolate and/or oblate configurations, are shown. The lower panel displays states that exhibit significant contributions from both prolate and oblate configurations, whereas the states shown in the upper panel are predominantly characterized by either prolate or oblate configurations. See text for more details.}
\label{fig:densities}
\end{figure*}

Figure \ref{fig:waves} displays the amplitudes of collective wave functions squared $|g(\beta_2, \beta_3)|^2$ for the lowest $0^+$ and $2^+$ levels of $^{12}$C. Even though they are not observables, these amplitudes provide a measure of quadrupole and octupole shape fluctuations in collective states. Moreover, they can be used to calculate expectation values of the deformation parameters $\beta_2$ and $\beta_3$ for each collective state:

\begin{subequations}
\begin{align}
\braket{\beta_2}_{\alpha}^{J;NZ;\pi} &= \sum_i |g_{\alpha}^{J;NZ;\pi}(q_i)|^2 {\beta_2}_i, \label{eq:avbeta2} \\
\braket{|\beta_3|}_{\alpha}^{J;NZ;\pi} &= \sum_i |g_{\alpha}^{J;NZ;\pi}(q_i)|^2 |{\beta_3}_i|, \label{eq:avbeta3}
\end{align}
\end{subequations}

\noindent where in the octupole direction we have taken the absolute value of deformation parameter since $\braket{\beta_3}_{\alpha}^{J;NZ;\pi}$ vanishes identically for all states with good parity. The values of $\braket{\beta_2}_{\alpha}^{J;NZ;\pi}$ and $\braket{|\beta_3|}_{\alpha}^{J;NZ;\pi}$ determine the dominant mean-field configurations in a collective state, and thereby enable the characterization of the corresponding intrinsic density. In Figure \ref{fig:densities} we plot the characteristic intrinsic nucleon densities of the first three $0^+$ and $2^+$ states of $^{12}$C. For each state the corresponding prolate and oblate deformation parameters $(\beta_2, \beta_3)$, shown in parenthesis, are calculated by averaging over the prolate and oblate configurations separately [cf. Eqs. (\ref{eq:avbeta2}) and (\ref{eq:avbeta3})]. For the average prolate or oblate $(\beta_2, \beta_3)$ we plot the corresponding intrinsic total nucleon density in the $xz$ plane. These densities are obtained by axial RHB calculations constrained to the average $(\beta_2, \beta_3)$ values. In each panel we also include the percentage of prolate or oblate configurations in the collective wave function. 

\indent Only the $0_1^+$ and $0_2^+$ states exhibit significant $(>20 \%)$ contributions from both prolate and oblate configurations, while other states predominantly correspond to either prolate or oblate shapes. In particular, the situation for the $0_1^+$ state is rather similar to that observed in the symmetry-restored $J^{\pi}=0^+$ energy curve of Fig. \ref{fig:PROJ1D}, where the prolate-deformed local minimum coexists with the oblate-deformed absolute minimum. In fact, the maximum of the $0_1^+$ collective wave function is also found at the oblate-deformed $(\beta_2, \beta_3) = (-0.4, 0.0)$ configuration. However, $|g(\beta_2, \beta_3)|^2$ has non-negligible contributions from configurations in a rather wide range of deformations: $\beta_2 \in [-1.2, 1.2]$ and $|\beta_3| \in [0.0, 1.4]$. Averaging over all these configurations accumulates significant contributions from prolate configurations that balance the influence of the oblate maximum, and would ultimately yield the nearly-spherical density distribution. This clearly does not reflect the actual physical picture of the ground state of $^{12}$C. In contrast, the two plots in the lower panel of Fig. \ref{fig:densities} reveal the complex structure of the $^{12}$C ground state. The collective wave function of the $2_1^+$ state is predominantly spread over ($\beta_2 < 0$, $\beta_3$) configurations and the corresponding density distribution is oblate-deformed. This could already be predicted from the low-energy oblate configurations in the $J^{\pi} = 2^+$ energy maps of Figs. \ref{fig:PROJ} and \ref{fig:PROJ1D}. Moreover, the average deformation $(\beta_2, \beta_3) = (-0.50, 0.58$) of the $4_1^+$ confirms the oblate nature of the ground-state band. Note that the present model is not suitable to investigate the role of triaxiality in \textit{yrast} states that was recently inferred experimentally \cite{lambarri14} and further suggested by some theoretical calculations (see e.g. Ref. \cite{tohsaki17} and references cited therein).

      The amplitude of the collective wave function of the $0_2^+$ state exhibits two maxima: one at small prolate deformations and the other one at larger oblate deformations. The corresponding density distributions are displayed in the lower right panel of Fig. \ref{fig:densities}. Of course, to obtain the expected triangular distribution of the Hoyle state \cite{tohsaki17} one needs to break axial symmetry, which has not been possible in the present work. Finally, the $K^{\pi} = 0_3^+$ band originates from the shoulder at large prolate $\beta_2$ values in Figs. \ref{fig:PROJ} and \ref{fig:PROJ1D}. In the intrinsic frame of reference, a large quadrupole deformation translates into a pronounced linear structure of the $0_3^+$ and $2_3^+$ collective states, as shown in the upper right panel of Fig. \ref{fig:densities}. A homogeneous alignment of $3\alpha$ particles would be described by a reflection-symmetric configuration in the intrinsic frame, that is, the corresponding octupole deformation would be negligible. Even though the maxima of the $0_3^+$ and $2_3^+$ amplitudes are indeed found at $\beta_3 = 0$ [see Fig. \ref{fig:waves}], fluctuations in the octupole direction are rather pronounced for both states. Consequently, the reflection-asymmetric $^{8}$Be$+\alpha$-like structure is observed in the intrinsic frame. The formation of linear chain structures in $^{12}$C was previously predicted by microscopic models \cite{kanada07,neff04}, but they are yet to be confirmed experimentally. Another interesting feature of these chains is their alleged susceptibility to bending, which would eventually lead to the formation of bent arm structures \cite{kanada07,neff04,ren18}. However, besides breaking reflection symmetry, such structures also break axial symmetry and they are therefore out of the scope of the present study.

\subsection{Electron-Nucleus Scattering Form Factors}

\begin{figure*}[!t]
   {\includegraphics[width=0.497\textwidth]{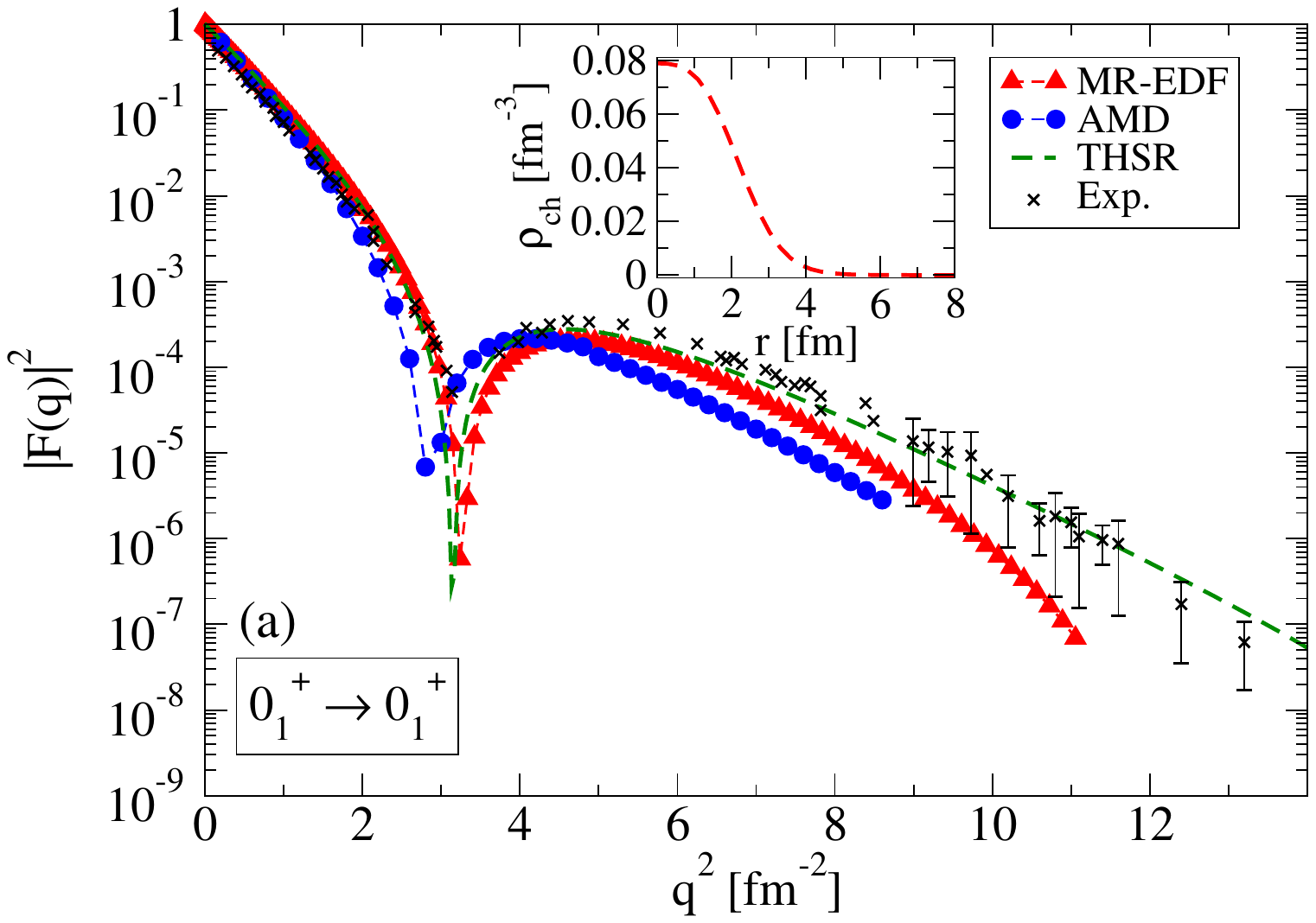}
      \includegraphics[width=0.497\textwidth]{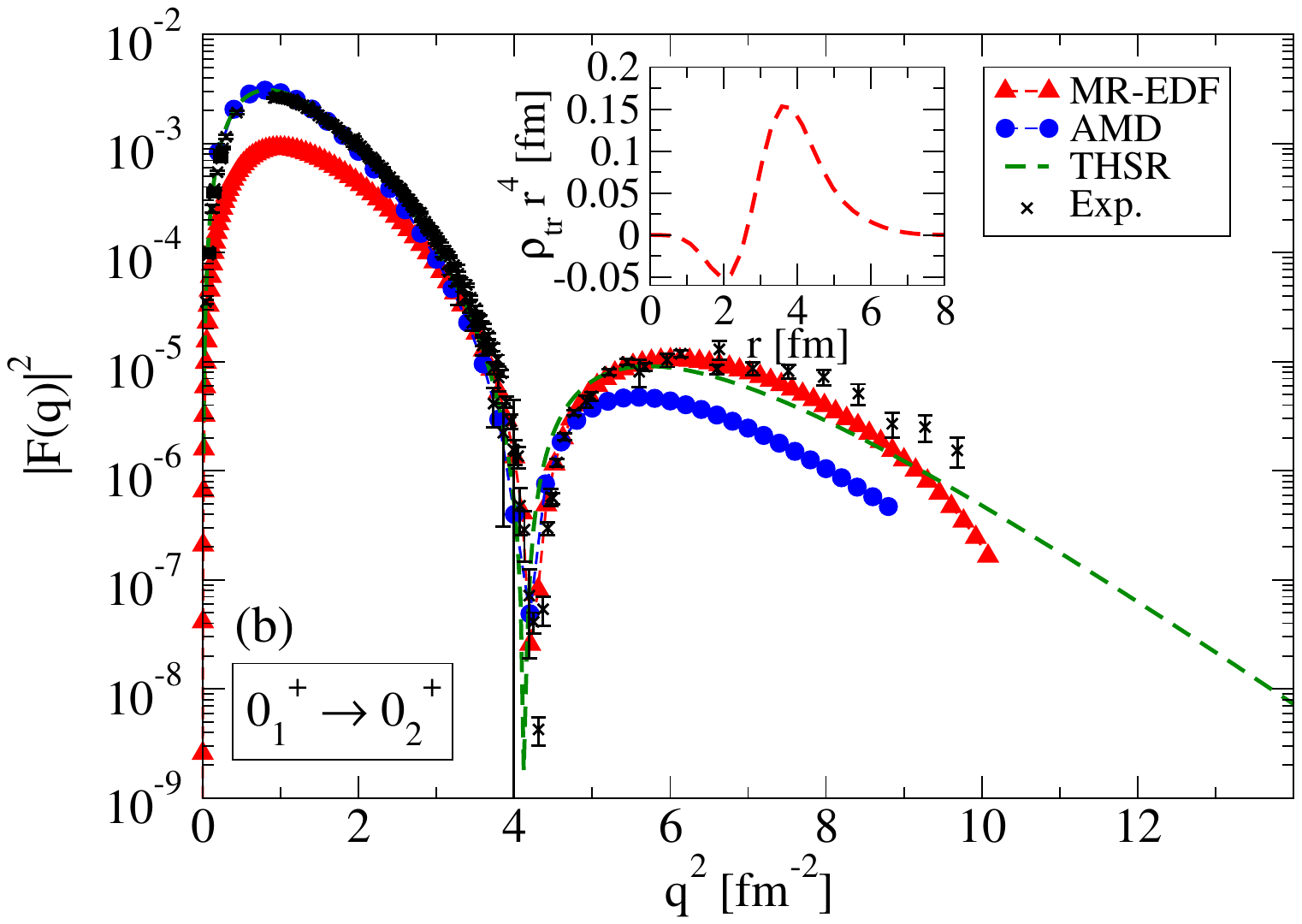}
      }
     \caption{Form factors for electron scattering on $^{12}$C for the $0_1^+ \rightarrow 0_1^+$ (left panel (a)) and $0_1^+ \rightarrow 0_2^+$ (right panel (b)) transitions. The results obtained in the present study (red triangles) are compared to available data for the elastic \cite{expelastic} and inelastic \cite{chernykh10} form factors, as well as with predictions of the AMD \cite{kanada07} and THSR \cite{funaki06} models. In addition, the insets show the corresponding charge density (left panel (a)) and the transition charge density (right panel (b)).}    
    \label{fig:formfactors}
\end{figure*}

Additional insight into the structure of collective states is provided by the form factors for electron-nucleus scattering. The formalism for computing these quantities within the MR-EDF framework was derived only recently \cite{yao15} and, in this section, we will calculate form factors for electron scattering on $^{12}$C for $0_1^+ \rightarrow 0_1^+$ (elastic) and $0_1^+ \rightarrow 0_2^+$ (inelastic) transitions. The basic ingredients of the calculation are the collective wave functions, whose amplitudes are shown in Fig. \ref{fig:waves}. As noted earlier, the wave functions of $0_1^+$ and $0_2^+$ are concentrated in the segment:  $-1.2 \le \beta_2 \le 1.2$ and $0 \le |\beta_3| \le 1.4$. Therefore, to reduce the computational task but without neglecting any physical content, only configurations contained within that interval of the  $\beta_2$ - $\beta_3$ plane are included in the calculation of form factors. In the left panel of Figure \ref{fig:formfactors} we display the calculated form factors $|F_0(q)|^2$ [cf. Eq. (\ref{eq:formfactor})] for elastic $0_1^+ \rightarrow 0_1^+$ scattering in comparison to the available experimental values. In addition, the inset shows the corresponding charge density, calculated as $\rho_{\text{ch}}(r) = \rho^{01}_{01,0}(r)/\sqrt{4\pi}$, where $\rho^{01}_{01,0}(r)$ is the diagonal element of the reduced transition density that enters into the calculation of the form factor [cf. Eqs. (\ref{eq:formfactor}) and (\ref{eq:redtrans})]. The present results are also compared with the predictions of the AMD model, and the THSR wave function model. AMD calculations \cite{kanada07} consider single nucleons as relevant degrees of freedom, and describe them in terms of Gaussian wave packets. On the other hand, the THSR framework is explicitly built as an $\alpha$-cluster model, that is, the relevant degrees of freedom are $\alpha$-particles in a Bose-condensed state \cite{funaki06}. In the low momentum transfer region ($q^2 < 2$ fm$^{ˆ'2}$) all three models predict similar results for the elastic form factor. However, with increasing values of the momentum transfer differences between the three curves becomes more pronounced. In particular, the first zero of $|F_0(q)|^2$ is found at approximately $q^2 \approx 3$ fm$^{ˆ'2}$ both in the present and THSR calculations, while the AMD model predicts this zero at a somewhat smaller value of the momentum transfer. Details of elastic form factors can be traced back to the properties of the corresponding charge distribution \cite{friedrich82,friedrich86}. The shift of the position of the first zero towards smaller values of $q^2$, in particular, can be attributed to the larger spatial extension of the charge density. Furthermore, the amplitude of the first maximum of $|F_0(q)|^2$ is related to the surface thickness of the charge distribution. Larger values of the surface thickness correspond to smaller amplitudes at the first maximum, and vice versa. We note that the experimental position and amplitude of the first maximum of $|F_0(q)|^2$ are reproduced by all three models. At very large values of $q^2$ the form factor calculated with the THSR model is in best agreement with experiment, whereas results obtained in the present study underestimate the experimental values. A similar trend was noted in Ref. \cite{yao15} for the case of $^{24}$Mg, and in Ref. \cite{fukuoka13} for $^{12}$C, where it was argued that the spreading of collective wave functions over many deformations generates a large smoothing of the one-body density and thus decreases the weights of large-momentum components of the charge density. 

	Because of its short lifetime, the structure of the Hoyle state can be probed by inelastic scattering from the ground state. In the right panel of Fig. \ref{fig:formfactors} we display the calculated form factor for the $0_1^+ \rightarrow 0_2^+$ transition in comparison with the available data, and predictions of the AMD and THSR models. The present calculation reproduces the position of the first maximum, even though the corresponding amplitude is underestimated in comparison to both experiment and theoretical results  obtained with the other two models. The position of the first zero is accurately reproduced by all three models. In addition, our model displays good agreement with experiment up to rather large $q^2$ values. The inset in the right panel of Fig. \ref{fig:formfactors} shows the corresponding transition charge density, $\rho_{\text{tr}}(r) = \rho^{02}_{01,0}(r)/\sqrt{4\pi}$, where $\rho^{02}_{01,0}(r)$ is the non-diagonal element of the reduced transition density that enters the calculation of form factor [see Eqs. (\ref{eq:formfactor}) and (\ref{eq:redtrans})]. The result of the present analysis for $\rho_{\text{tr}}(r) r^4$ can be directly compared to the experimental transition charge density that corresponds to the form factor of Fig. \ref{fig:formfactors}, as well as to the predictions of the FMD model and the $\alpha$-cluster model (cf Fig. 3 of Ref. \cite{chernykh10}). While the position of the minimum of $\rho_{\text{tr}}(r) r^4$ is very similar for all four curves considered, the present calculation predicts a somewhat weaker amplitude in comparison to both the experiment and the other models. Furthermore, the FMD and $\alpha$-cluster models overestimate the experimental maximum value of the  $\rho_{\text{tr}}(r) r^4$ curve, located at $r \approx 4$ fm. Our calculation, on the other hand, notably underestimates this value. This difference is then naturally reflected in the lower value of the form factor compared to experiment, particularly at low values of momentum transfer. The inclusion of the triaxial degree of freedom, that undeniably plays an important role in the $0_2^+$ state and that could also influence the structure of the ground state, would likely modify the calculated transition charge density and, consequently, the corresponding form factor. Whether such an extension of the model space actually leads to results that are closer to experimental values, remains to be examined in future studies.


\section{Summary}
\label{sec:conclusion}

The low-lying excitation spectrum and cluster structures in $^{12}$C isotope have been analyzed using a ``beyond mean-field'' approach based on global energy density functionals. Axially-symmetric and reflection-asymmetric RHB states extending over a wide range of quadrupole and octupole deformations have first been projected onto good angular momentum, particle number, and parity, and subsequently configuration mixing implemented using the generator coordinate method. Although we have not explicitly included triaxial shapes, that play an important role in the $K^{\pi}=0_2^+$ band and whose role in $K^{\pi}=0_1^+$ and $K^{\pi}=0_3^+$ bands is yet to be elucidated, the present model successfully reproduces many spectroscopic features of $^{12}$C. In particular, empirical properties of the ground-state band have been accurately reproduced, including available data on excitation energies and spectroscopic quadrupole moments, as well as the intraband quadrupole transition strengths. The rotational band built on the state $K^{\pi}=0_3^+$ corresponds to a strongly prolate deformed shape characterized by a linear $\alpha$-chain structure in the intrinsic frame. Notably, the calculated $0_2^+$ state is located only about $800$ keV above its experimental excitation energy. In addition, both the elastic and inelastic form factors are in good agreement with the experimental values for a rather wide interval of momentum transfer. In this particular aspect, the accuracy of the present global approach can compete with that of the most successful microscopic cluster models.

The symmetry-conserving global method used in the present study is based on the universal framework of energy density functionals. In particular, starting from the functional DD-PC1 that was exclusively adjusted to the experimental binding energies of a set of 64 deformed nuclei in the mass regions $A \approx 150 -180$ and $A \approx 230 - 250$, and using a basis and method that do not \textit{a priori} assume the existence of cluster structures, this model enables a consistent, parameter-free calculation of collective excitation spectra and the corresponding electric transition strengths in light nuclear systems such as $^{12}$C. Of course, such a global approach may not be able to describe all the details of excited states configurations, that often cannot be reproduced even by models specifically designed and fine-tuned to this mass region. The EDF-based  framework, however, especially when extended to take into account ''beyond mean-field'' correlations related to multiple broken symmetries and quantum fluctuations, presents one of the most promising theoretical methods for a unified description of complex quantum-liquid and cluster aspects of atomic nuclei.

\newpage
\acknowledgements{
This work was supported in part by 
the Croatian Science Foundation -- project "Structure and Dynamics
of Exotic Femtosystems" (IP-2014-09-9159) and the QuantiXLie Centre of Excellence,
a project cofinanced by the Croatian Government and European Union through the 
European Regional Development Fund - the Competitiveness and Cohesion Operational
Programme (KK.01.1.1.01).
}


\newpage

\begin{thebibliography}{99}
\bibitem{beck10} C. Beck, Ed., Clusters in Nuclei, Vol. 1, Lecture Notes in Physics 818 (Springer-Verlag, Berlin, Heidelberg, 2010).
\bibitem{beck12} C. Beck, Ed., Clusters in Nuclei, Vol. 2, Lecture Notes in Physics 848 (Springer-Verlag, Berlin, Heidelberg, 2012).
\bibitem{beck14} C. Beck, Ed., Clusters in Nuclei, Vol. 3, Lecture Notes in Physics 875 (Springer-Verlag, Berlin, Heidelberg, 2014).
\bibitem{horiuchi12} H. Horiuchi, K. Ikeda, and K. Kato, Prog. Theor. Phys. Suppl. 192, 1 (2012).
\bibitem{freer17} M. Freer, H. Horiuchi, Y. Kanada-En'yo, D. Lee, and Ulf-G. Mei{\ss}ner, arXiv:1705.06192v1 [nucl-th].
\bibitem{kanada07} Y. Kanada-En'yo, Prog.  Theor.  Phys. 117, 655 (2007).
\bibitem{neff04} T. Neff and H. Feldmeier, Nucl. Phys. A 738, 357 (2004).
\bibitem{chernykh07} M. Chernykh, H. Feldmeier, T. Neff, P. von Neumann-Cosel, and A. Richter, Phys. Rev. Lett 98, 032501 (2007).
\bibitem{tohsaki01} A. Tohsaki, H. Horiuchi, P. Schuck, and G. R\"{o}pke, Phys. Rev. Lett. 87, 192501 (2001).
\bibitem{funaki15} Y. Funaki, Phys. Rev. C 92, 021302 (2015).
\bibitem{fukuoka13} Y. Fukuoka, S. Shinohara, Y. Funaki, T. Nakatsukasa, and K. Yabana, Phys. Rev. C 88, 014321 (2013).
\bibitem{ren18} Z. X. Ren, S. Q. Zhang, P. W. Zhao, N. Itagaki, J. A. Maruhn, and J. Meng, arXiv:1805.07901 [nucl-th].
\bibitem{tohsaki17} A. Tohsaki, H. Horiuchi, P. Schuck, and G. R\"opke,
Rev. Mod. Phys. 89, 011002 (2017).
\bibitem{kumar18} M. Kumar Raju, Phys. Lett. B 777, 250 (2018).
\bibitem{lambarri14} D. J. Mar\'in-L\'ambarri, R. Bijker, M. Freer, M. Gai, Tz. Kokalova, D. J. Parker, and C. Wheldon, Phys. Rev. Lett. 113, 012502 (2014).
\bibitem{dellaquila17} D. Dell'Aquila \emph{et al.}, Phys. Rev. Lett. 119, 132501 (2017).
\bibitem{smith17} R. Smith \emph{et al.}, Phys. Rev. Lett. 119, 132502 (2017).
\bibitem{VALR.05} D. Vretenar, A. V. Afanasjev, G. A. Lalazissis, and P. Ring, Phys. Rep. {409}, 101 (2005).
\bibitem{niksic11} T. Nik\v{s}i\'c, D. Vretenar, and P. Ring, Prog. Part. Nucl. Phys. 66, 519 (2011).
\bibitem{arumugam05} P. Arumugam, B. K. Sharma, and S. K. Patra, Phys. Rev. C 71, 064308 (2005).
\bibitem{maruhn06} J. A. Maruhn \emph{et al.}, Phys. Rev. C 74, 044311 (2006).
\bibitem{ebran12} J.-P. Ebran, E. Khan, T. Nik\v{s}i\'c, and D. Vretenar, Nature 487, 341 (2012).
\bibitem{ebran14a} J.-P. Ebran, E. Khan, T. Nik\v{s}i\'c and D. Vretenar, Phys. Rev. C 89, 031303(R) (2014).
\bibitem{ebran14b} J.-P. Ebran, E. Khan, T. Nik\v{s}i\'c and D. Vretenar, Phys. Rev. C 90, 054329 (2014).
\bibitem{ebran18} J.-P. Ebran, E. Khan, R.-D. Lasseri, and D. Vretenar, Phys. Rev. C 97, 061301(R) (2018).
\bibitem{zhou16} E. F. Zhou, J.M. Yao, Z.P. Li, J. Meng, and P. Ring, Phys. Lett. B 753, 227 (2016).
\bibitem{marevic18} P. Marevi\'c, J.-P. Ebran, E. Khan, T. Nik\v{s}i\'c, and D. Vretenar, Phys. Rev. C 97, 024334 (2018).
\bibitem{bender03} M. Bender, P.-H. Heenen, and P.-G. Reinhard, Rev. Mod. Phys. 75, 121 (2003).
\bibitem{egido16} J.L. Egido, Phys. Scr. 91, 073003 (2016).
\bibitem{RS.80} P. Ring, P. Schuck, The Nuclear Many Body Problem, Springer-Verlag Berlin Heidelberg, 1980. 
\bibitem{varshalovich88} D.A. Varshalovich, A.N. Moskalev, and V.K. Khersonskii, Quantum Theory of Angular Momentum (World Scientific, Singapore, 1988).
\bibitem{niksic08} T. Nik\v{s}i\'c, D. Vretenar, and P. Ring, Phys. Rev. C 78, 034318 (2008).
\bibitem{duguet04} T. Duguet, Phys. Rev. C 69, 054317 (2004).
\bibitem{tian09} Y. Tian, Z. Y. Ma, and P. Ring, Phys. Lett. B 676, 44 (2009).
\bibitem{gambhir90} Y.K. Gambhir, P. Ring, and A. Thimet, Ann. Phys. (N.Y.) 198, 132 (1990).
\bibitem{fomenko70} V. N. Fomenko, J. Phys. (GB) A3, 8 (1970).
\bibitem{hwg} D.L. Hill and J.A. Wheeler, Phys. Rev. 89, 1102 (1953); J.J. Griffin and J.A. Wheeler, Phys. Rev. 108, 311 (1957).
\bibitem{robledo10}  L. M. Robledo, J. Phys. G: Nucl. Part. Phys. 37, 064020 (2010).
\bibitem{rodriguez02} R. Rodr\'iguez-Guzm\'an, J. L. Egido, and L. M. Robledo, Nucl. Phys. A 709, 201 (2002).
\bibitem{yao15} J. M. Yao, M. Bender, and P.-H. Heenen, Phys. Rev. C 91, 024301 (2015).
\bibitem{brown83} B. A. Brown, R. Radhi, and B. H. Wildenthal, Phys. Rep. 101, 313 (1983).
\bibitem{kelley17} J. H. Kelley, J. E. Purcell, and C.G. Sheu, Nucl. Phys. A 968, 71-253 (2017).
\bibitem{nndc} Brookhaven National Laboratory NNDC National Nuclear Data Center.
http://www.nndc.bnl.gov. F. Ajzenberg-Selove, Nucl. Phys. A 506, 1-158 (1990).
\bibitem{navratil07} P. Navr\'atil, V. G. Gueorguiev, J. P. Vary, W. E. Ormand, and A. Nogga, Phys. Rev. Lett 99, 042501 (2007).
\bibitem{neff12} T. Neff, J. Phys.: Conf. Ser. 403, 012028 (2012).
\bibitem{uegaki77} E. Uegaki, S. Okabe, Y. Abe, and H. Tanaka, Prog. Theor. Phys. 57, 1262 (1977).
\bibitem{expelastic} I. Sick and J. S. McCarthy, Nucl. Phys. A 150, 631-654  (1970). A. Nakada, Y. Torizuka, and Y. Horikawa, Phys. Rev. Lett. 27, 745 (1971). P. Strehl and T. H. Schucan, Phys. Lett. B 27, 641-643 (1968).
\bibitem{chernykh10} M. Chernykh, H. Feldmeier, T. Neff, P. von Neumann-Cosel, and A. Richter, Phys. Rev. Lett 105, 022501 (2010).
\bibitem{funaki06} Y. Funaki, A. Tohsaki, H. Horiuchi, P. Schuck, and G. R\"{o}pke, Eur. Phys. J. A 28, 259–263 (2006).
\bibitem{amado80} R. D. Amado, J. P. Dedonder, and F. Lenz, Phys. Rev. C 21, 647 (1980).
\bibitem{friedrich82} J. Friedrich and N. Voegler, Nucl. Phys. A 373, 192-224 (1982).
\bibitem{friedrich86} J. Friedrich, N. Voegler, and P.-G. Reinhard, Nucl. Phys. A 459, 10-34 (1986).

\end{thebibliography}
\end{document}